# First-principle prediction of the existence of $C_{64}$-graphyne and its nitrogen and boron substitutions

Hui Li[a], Zihua Xin[a], Junxian Liu[a], Jiali Wu[a], M. Yu[b]

[a] *Department of Physics, Shanghai University, Shanghai 200444, China*

[b] *Department of Physics and Astronomy, University of Louisville, Louisville, KY, 40292, USA*

**Abstract**

By using of the first-principles calculations based on density functional theory, a novel monolayer planar structure named $C_{64}$-graphyne is predicted. Tetratomic and hexatomic rings, as well as C-C triple bonds exist in this new stable structure with the lattice parameter of 9.291 Å. The carbon hexatomic ring in $C_{64}$-graphyne contains two quite distinct C-C bonds, which is known as cyclohexatriene. Its electronic band structures show a semiconductor nature with a narrow direct band gap of 0.35 eV. Interestingly, by substituting one nitrogen atom for a carbon atom in the hexatomic ring or a sp hybridized carbon atom on the carbon chain of $C_{64}$-graphyne, two stable planar structures are obtained. Such doping with nitrogen atom induced metal properties in this monolayer structure. Further investigation shows that the alternating substitution of boron and nitrogen atoms for all carbon atoms in $C_{64}$-graphyne also induced a new stable structure, the $(BN)_{64}$ structure. The electronic studies proved its insulator property with the band gap of 4.08 eV.

**Key words:** First-principle calculation, Phonon dispersion, Electronic structure, Substitution and doping

# 1. Introduction

Various allotropes of carbon have always been the focus of the research. Graphene has become a new research hotspot since its successful synthesis in 2004 [1], which is expected to replace the original silicon-based semiconductor. More and more people began to investigate graphene, such as electronic structure [2], optical transparency [3] and so on. However, due to the fact that graphene is a zero-bandgap semiconductor, the use of graphene in semiconductor device has been limited, such as the field-effect tube (FET) cannot be made. Therefore, researchers began to modulate their band gap by dividing the plane structure into nano band structures [4] or doping [5]. The studies show that the doping of boron, nitrogen, oxygen and fluorine atoms into graphene can effectively change the bandgap, and the doping of fluorine atoms will further induce magnetic properties of graphene [5]. At present, the structure of boron and nitrogen doped graphene has been successfully synthesized [6–8]. Hence, it is expectable to find more carbon allotropes with significant performance.

The two-dimensional carbon materials containing acetylenic linkages are often called graphyne, which was first proposed theoretically by Baughman *et al*. [9] in 1987. In 2009, graphyne (i.e., graphdiyne) was synthesized by Li *et al*. [10], on the surface of copper via across-coupling reaction using hexaethynylbenzene. The successful synthesis of graphyne make it possible to apply this material to new semiconductor, and this stimulated the research enthusiasm of graphyne. Since then, there have been many theoretical studies on graphyne and its derivatives. Zhang *et al*. [11] found that the presence of the acetylenic linkages in graphynes leads to a significant change in fracture stress and Young's modulus with the degree of change being proportional to the percentage of the linkages. Long *et al*. [12] found that graphyne has high electron mobility, and with a band gap of 0.45eV. The γ-graphyne and graphdiyne have good thermal stability [13]. Studies by zhang [14] and liu *et al*. [15]. show that the doping of germanium atoms can effectively adjust the bandgaps of α-graphyne, γ-graphyne and their related structures. The structure stability of silicon-doped graphdiyne can be maintained within a wide range of temperature [16], it also reported that a four-membered ring which contains both carbon and silicon begins to appear beside the six-membered ring when a certain higher temperature is reached, furthermore, the carbon-silicon four-membered ring will not disappear when the temperature drops back to zero. In addition, the doping of other atoms such as boron [17–19], nitrogen [17,18], oxygen [18] and aluminum [19] can effectively regulate the bandgap of graphyne. By doping, the hydrogen storage capacity of graphyne can be improved effectively [20–24]. In addition, the graphyne with the building units of parallelogram also entered the vision of researchers in 2015 [25]. It is a monolayer planar material with Dirac cones.

Song *et al.* [26] studied a new allotrope of carbon which contains both tetratomic and hexatomic rings, called graphenylene. It is a special carbon structure with cyclohexatriene unit. There are two distinct C-C bonds in the unit, with bond lengths of 1.366 Å and 1.467 Å, respectively. The results of calculations show that the structure with cyclohexatriene unit is more stable than γ-graphyne and carbyne. When graphenylene was divided into ribbon structure, the band gap can be increased from 0.025 eV to 1 eV evidently. Another four carbon allotropes have been studied by Lu *et al.* [27], $C_{65}$-, $C_{63}$-, $C_{31}$- and $C_{41}$-sheets are consist of six-membered and five-membered rings, six-membered and three-membered rings, three-membered ring and one carbon atom, four-membered ring and one carbon atom, respectively. The electronic structure studies of these four stable structures show that they are all metallic. The hydrogen storage capacity of these structures can be greatly improved when adsorb lithium atoms at different positions [28]. Previous studies have shown that the basic units that make up graphyne are diverse. The diversity of such units and connections between the units give rise to the diversity of graphyne. It is reasonable to believe the existence of new graphynes with excellent properties. Therefore, to constantly enrich the family member of graphyne and its derivatives has been a common goal of many scientific researchers. The purpose of this paper is to find new graphyne with specific properties by first-principles calculations. We predicted four types of stable monolayer planar structures through simulation. We will report our results as follows: The calculation method is described in detail in Section 2. The geometric structures, phonon dispersion spectrum, charge distribution and electronic structures of these four predicted structures are calculated and discussed in Section 3. Finally, a brief conclusion is given in Section 4.

**2. Computational methods**

All the calculations of structural and electronic properties are based on the density functional theory (DFT) with the projector-augmented-wave (PAW) method in the Vienna ab initio Simulation Package (VASP). The exchange correlation potential is evaluated by Perdew-Burke-Ernzerh of (PBE) functional using the generalized gradient approximation (GGA). The cutoff energy of plane wave is 550 eV. A 11×11×1 Monkhorst-pack k-point mash was used to sample the Brillouin zone. A vacuum layer of 20 Å was applied perpendicular to the sheets to avoid the interlayer interactions. In all these calculations the convergence criteria for total energy in the selfconsistent field iteration was set to be $10^{-6}$ eV and the optimizations have been carried out by keeping the volume of the unit cell constant until the Hellmann-Feynman force component on each atom is less than $10^{-3}$ eV/Å. The vibrational properties are calculated by using PHONOPY code, which can directly use the force constants calculated by density functional perturbation theory (DFPT) as implemented in the VASP code. Here, 2×2×1 supercells are considered for all structures.

**3. Results and discussion**

*3.1. Structural optimization and characterization*

The first stable structure we obtained is a structure which contains carbon tetratomic rings and hexatomic rings, as shown in Fig. 1(a). The simulation reveals that this is a carbon monolayer plane containing *sp* hybridization, named $C_{64}$-graphyne. To examine the stability of the doped structures with B or N atoms in present system, a large number of optimization are performed. The results show that when C atom is substituted by B atoms, the structure will be destroyed. When a single N atom replaces one C atom, two stable stuctures are obtained. In one structure, the C atom in hexatomic ring is substituted with N atom. In the other structure, the C atom in the chain connecting two tetratomic rings is substituted with N atom. We named them as $N_{ring}$C-graphyne and $N_{chain}$C-graphyne, respectively. When substitute all C atoms with B and N atoms alternately, a new stable structure is also obtained, we named it as $(BN)_{64}$ structure. The $N_{ring}$C-graphyne, the $N_{chain}$C-graphyne and the $(BN)_{64}$ structure are shown in Fig. 1(b, c, d), respectively. The lattice parameters of these four new structures are given in Table 1. Compared with $C_{64}$-graphyne, the lattice constants of $N_{chain}$C-graphyne and $N_{ring}$C-graphyne are smaller, while the lattice constant of $(BN)_{64}$ structure is relatively large. In addition, since the total energy of $N_{chain}$C-graphyne is lower than that of $N_{ring}$C-graphyne, it is more stable for the structure in which N atoms to be doped in the chain instead of the ring.

In order to the understand hybrid characteristics of these four structures. We analyzed various bonds and bond angles of the four structures (which are labeled in Fig. 2), and listed their values in Table 2 and Table 3, respectively. The results show that in $C_{64}$-graphyne, two different C-C bonds exist alternately in the hexatomic ring of carbon, with the bond lengths of 1.378 Å (bonds 1, 3, 5 in Fig. 3($a_1$)) and 1.466 Å ( bonds 2, 4, 6 in Fig. 3($a_1$)), and all the bond angles in the rings are 120° ( bond angles $\alpha_1$-$\alpha_6$ in Fig. 3($a_2$)). It is different from the standard benzene ring in graphene and other graphyne. For example, the C-C bond lengths of hexatomic rings are 1.426 Å and 1.432 Å in γ-graphyne and graphdiyne, respectively, which are among the two bond lengths of the hexatomic rings in $C_{64}$-graphyne. This hexatomic ring of carbon consisting of two different bond lengths is known as the cycloheptatriene unit [26]. In addition, the length of the bond 7 and bond 9 in the tetratomic ring are 1.459 Å, and the bond 8 is 1.508 Å, which makes the tetratomic ring behaved as regular isosceles trapezoidal structures. The bond length of C-C bond connecting the two tetratomic rings (bond 11 in Fig. 3($a_1$)) is 1.253 Å, indicating that bond 11 is an acetylene linkage between two sp-hybridized carbon atoms. On the other hand, the bond angles of $\alpha_{12}$ and $\alpha_{13}$ connected to the anacetylene linkage are both 170°13′.

When a nitrogen atom is doped into the hexatomic ring of $C_{64}$-Graphyne, the hexatomic ring of $N_{ring}$C-graphyne is still not a standard benzene ring, with long and short bond alternately. However, there are a little difference among the three long bonds, so are the three short bonds. Therefore, the ring is a quasi-cyclictriltene unit, in which the lengths of bonds 1, 3 and 5 are shorter, bonds 2, 4, 6 are longer. Due to the substitution of nitrogen atoms, bond angles in the hexatomic ring is slightly deviated from the original 120°. We can see the value of bond angles in Table 3 where the tetratomic ring is no longer a trapezoid structure.

The hexatomic ring in the $N_{chain}$C-graphyne is also a distorted quasi-cyclictriltene unit, and the tetratomic ring is no longer a trapezoid structure. The length of bonds 10 and 11 located on the chain is significantly shorter than that of $C_{64}$-graphyne. The C-N bond is the shortest with the length of 1.202 Å, indicating that the two atoms of C and N at the C-N bond in the stable structure is *sp*-hybridized. The bond angles $α_{12}$ and $α_{13}$ are all changed compared with that in $C_{64}$-graphyne.

In $(BN)_{64}$ structure, all the length of B-N bonds are longer than the corresponding bonds in $C_{64}$-graphyne, resulting in the lattice constants of $(BN)_{64}$ structure larger than the other three structures. Obviously the ring is similar to a cyclictriltene unit, with bond angles of 121°50′and 118°10′. Bonds 1, 3, 5 in the hexatomic ring are short with the length of 1.405 Å, and bonds 2, 4, 6 are long with the length of 1.484 Å. The tetratomic ring is no longer a trapezoid structure. The length of bond 11 is 1.281 Å, which shows anacetylene linkage property formed by B and N atoms at *sp*-hybridized states.

Moreover, as the four structures contain a variety of pores with different size, The largest pore occures in $(BN)_{64}$ structure consists of 18 atoms, it is worth study in separating big size molecules.

*3.2. Phonon dispersion*

To verify the dynamics stability of the four structures, the phonon spectrums were studied, and the results are presented in Fig. 4. It is found that the phonon frequencies of $C_{64}$-graphyne, $N_{ring}$C-graphyne, $N_{chain}$C-graphyne and $(BN)_{64}$ structure are all positive. The presence of *sp* hybridization in those structures induced the existence of high-frequency vibration modes occurred between 60 THz and 70 THz [29]. This is similar to the previously predicted phonon spectra of α-, β-, γ- and 6,6,12-graphyne [30]. The result of phonon spectrum shows that the four structures are all stable and reliable.

*3.3. Charge distribution*

The electron localization function (ELF) was calculated to investigate the charge distribution around each atom in these stable structures. The charge distribution is given in Fig. 5. The scale value of 1 in the left column of Fig. 5 indicates that the electron density is high, corresponding to the completely localized states of the electron, while the value of 0.5 means the electron density is lower, corresponding to the delocalized states of the electrons, and the value of 0 indicates that there is no electron distribution.

It is found that the charge distribution around the C atoms in the hexatomic ring of $C_{64}$-graphyne (Fig. 5(a)) does not show the hexagonal symmetry, which is obviously different from that of the standard benzene ring. Furthermore, the charges density between the two C atoms in *sp* hybridized states in the chain is higher than that near other atoms. On the other hand, in $N_{ring}C$-graphyne and $N_{chain}C$-graphyne (Fig. 5(b, c)), the charge density near the N atom is larger than that near C atoms. In $(BN)_{64}$ structure (Fig. 5(d)), the charge density near the N atom is obviously larger than that near the B atom, which means the electrons of the N atoms are highly localized.

*3.4. Electronic structures*

Calculated the electronic structures of these monolayer planar structures are presented in Fig. 6. To compare the electronic structural characteristics of these structures, we presented the band gaps and their positions of $C_{64}$-graphyne, $N_{ring}C$-graphyne, $N_{chain}C$-graphyne and $(BN)_{64}$ structure in Table 4. The results show that $C_{64}$-graphyne is a direct bandgap semiconductor with the band gap of 0.35 eV. The valence band maximum and conduction band minimum locate at the M point in the Brillouin zone. The band near the Fermi level is relatively wide, which indicates that the electron effective mass in this band is smaller and the nonlocality is stronger. There is a relatively wide band going through the Fermi level in the band structures of $N_{ring}$C-graphyne and $N_{chain}$C-graphyne, indicating the metallic property. The $(BN)_{64}$ structure has a large band gap of 4.08eV, which means the $(BN)_{64}$ structure acts as an insulator. The band close to the Fermi level is especially narrow, indicating that the localization of electrons is quite strong and the effective mass of electrons is relatively large.

Detail analyses on contribution of the electrons in different orbit to the electronic structure of these structures have been carried out by calculating the total density of states (Total DOS), the local density of states (LDOS) and the partial density of states (PDOS) of the four structures.

As show in Fig. 7, the Total DOS and PDOS of $C_{64}$-graphyne show that the electron states near the top of valence band and the bottom of conduction band are completely contributed by the electrons in the $p_z$-orbital.

The LDOS of $N_{ring}$C-graphyne, $N_{chain}$C-graphyne (Fig. 8(a) and (b)), on the other hand, show that near the Fermi lever, the density of states of N atoms is almost zero, and the density of states is mainly contributed by C atoms. The electrons in the $s$-, $p_x$-and $p_y$-orbitals near the Fermi level have little contribution. Therefore, it can be concluded that the density of states near the Fermi level is mainly contributed by the electrons in the $p_z$-orbital of carbon atoms. It is similar to that of $C_{64}$-graphyne, indicating that the contribution from N atom is very small compared with the contribution from C atoms near the Fermi surface when substituting one C atom with one N atom.

The Total DOS of the $(BN)_{64}$ structure (Fig. 8(c)) shows sharp peaks near the top of valence band and the bottom of conduction band, indicating that the electrons are highly localized, and $(BN)_{64}$ structure is an insulator. It can be seen from the LDOS and PDOS that the bottom of conduction band is contributed by the electron in the $p_z$-orbital of the B atom, while the density at the top of the valence band is contributed by the electron in the $p_z$-orbital of the N atom.

## 4. Conclusions

We have systematically studied the geometric structures, phonon dispersion spectrum, charge distribution and electronic structures of four new stable monolayer systems using the first-principles calcuations.

As a novel graphyne, $C_{64}$-graphyne is a special structure which have both hexatomic and tetratomic rings of carbon. Very interestingly, the hexatomic ring in this structure contains two different C-C bonds which is the typical characteristics of cyclohexatriene, this is different from the traditional hexatomic ring of carbon. Moreover, different from in other graphynes, the distribution of electrons around the carbon atoms in the hexatomic of $C_{64}$-graphyne is not uniform. The tetratomic ring in $C_{64}$-graphyne is regular isosceles trapezoid structure. The bond length between the two *sp* hybridized carbon atoms is 1.253 Å, which shows acetylene linkage property. The electronic calculation shows that the $C_{64}$-graphyne is a semiconductor, with a band gap of 0.35 eV.

When substituting C atom with N atom either in ring or in chain, two stable monatomic layer structures were obtained, one is called $N_{ring}$C-graphyne and the other one is $N_{chain}$C-graphyne. Calculated results show that these two new sheets are exact plane structure. Their hexatomic rings are still similar to cyclohexatriene unit, where the lengths of the bonds are alternating, though the three long bonds are not exactly the same, and neither the three short bonds. Their tetratomic rings are distorted, and are no longer regular isosceles trapezoidal structures. The bond between the two *sp* hybridized C atoms is 1.250 Å and 1.202 Å in $N_{ring}$C-graphyne and $N_{chain}$C-graphyne, respectively. The doping of nitrogen atom makes both structures show metallic properties. The total energy shows that it will be more stable when the single nitrogen atom substitutes a carbon atom in the hexatomic ring rather than in the tetratomic ring.

One stable monatomic layer sheets is obtained by substituting carbon atoms with boron and nitrogen atoms alternately, the hexatomic ring in the $(BN)_{64}$ structure contains two kinds of bonds with the lengths of 1.484 Å and 1.405 Å. This has never been seen from the reported works. The tetratomic rings in the $(BN)_{64}$ structure are no longer regular isosceles trapezoidal structures.The electronic structures shows that this structure behave as insulator.

Since these newly discovered structures contain a variety of pores with different size, and the largest pore occures in $(BN)_{64}$ structure, it is worth study in separating big size molecules.

**Table 1**

Lattice constant and total energy of $C_{64}$-graphyne, $N_{ring}C$–graphyne, $N_{chain}C$-graphyne and $(BN)_{64}$ structure

| Item | $C_{64}$-graphyne | $N_{ring}C$-graphyne | $N_{chain}C$-graphyne | $(BN)_{64}$ |
|---|---|---|---|---|
| Lattice Constant(Å) | 9.291 | 9.236 | 9.256 | 9.417 |
| Total Energy(eV/cell) | -151.930 | -151.016 | -151.383 | -147.753 |
| Energy per atom(eV) | -8.441 | -8.390 | -8.410 | -8.209 |

**Table 2**

Bond lengths of $C_{64}$-graphyne, $N_{ring}C$-graphyne, $N_{chain}C$-graphyne and $(BN)_{64}$ structure

| Bond | $C_{64}$-graphyne (Å) | $N_{ring}C$-graphyne (Å) | $N_{chain}C$-graphyne (Å) | $(BN)_{64}$ (Å) |
|---|---|---|---|---|
| 1 | 1.378 | 1.328 | 1.367 | 1.405 |
| 2 | 1.466 | 1.465 | 1.481 | 1.484 |
| 3 | 1.378 | 1.368 | 1.367 | 1.405 |
| 4 | 1.466 | 1.474 | 1.474 | 1.484 |
| 5 | 1.378 | 1.362 | 1.374 | 1.405 |
| 6 | 1.466 | 1.463 | 1.478 | 1.484 |
| 7 | 1.459 | 1.448 | 1.463 | 1.468 |
| 8 | 1.508 | 1.470 | 1.475 | 1.529 |
| 9 | 1.459 | 1.432 | 1.458 | 1.469 |
| 10 | 1.343 | 1.351 | 1.323 | 1.373 |
| 11 | 1.253 | 1.250 | 1.202 | 1.281 |
| 12 | 1.343 | 1.358 | 1.356 | 1.366 |

**Table 3**

Bond angles of $C_{64}$-graphyne, $N_{ring}C$-graphyne, $N_{chain}C$-graphyne and $(BN)_{64}$ structure

| Angle | $C_{64}$-graphyne | $N_{ring}C$-graphyne | $N_{chain}C$-graphyne | $(BN)_{64}$ |
|---|---|---|---|---|
| 1 | 120° | 121° 09′ | 120° 45′ | 121° 50′ |
| 2 | 120° | 119° 25′ | 120° 43′ | 118° 10′ |
| 3 | 120° | 118° 34′ | 118° 42′ | 121° 50′ |
| 4 | 120° | 121° 37′ | 120° 43′ | 118° 10′ |
| 5 | 120° | 118° 06′ | 120° 33′ | 121° 50′ |
| 6 | 120° | 121° 09′ | 118° 34′ | 118° 10′ |
| 7 | 90° 50′ | 90° 24′ | 89° 47′ | 85° 24′ |
| 8 | 90° 50′ | 89° 52′ | 90° 4′ | 96° 21′ |
| 9 | 89° 10′ | 89° 29′ | 89° 45′ | 83° 50′ |
| 10 | 89° 10′ | 90° 15′ | 90° 24′ | 94° 25′ |
| 11 | 129° 47′ | 132° 27′ | 129° 53′ | 128° 52′ |
| 12 | 170° 13′ | 162° 28′ | 169° 18′ | 163° 06′ |
| 13 | 170° 13′ | 173° 07′ | 173° 47′ | 177° 57′ |

**Table 4**

The band gaps and their positions of $C_{64}$-graphyne, $N_{ring}C$-graphyne, $N_{chain}C$-graphyne and $(BN)_{64}$ structure.

| Item | $C_{64}$-graphyne | $N_{ring}C$-graphyne | $N_{chain}C$-graphyne | $(BN)_{64}$ |
|---|---|---|---|---|
| Gap position | M | —— | —— | M |
| Gap value(eV) | 0.35 | Metal | Metal | 4.08 |
| Gap type | Direct | —— | —— | Direct |

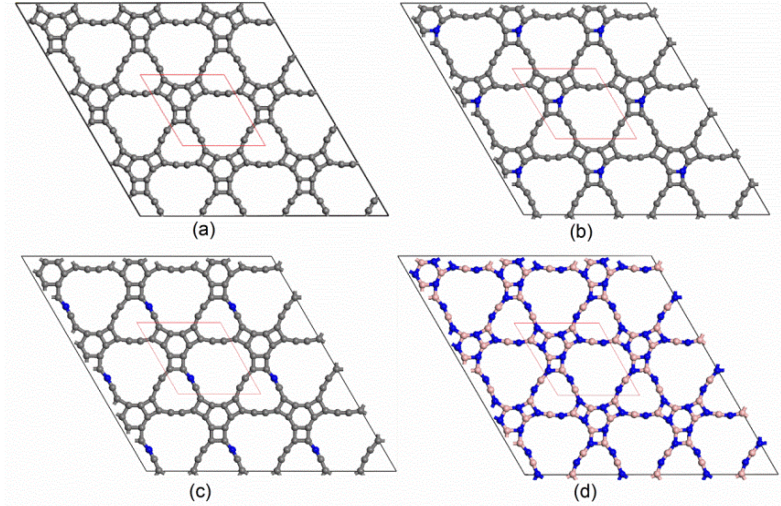

**Fig. 1.** Schematic structures of (a) $C_{64}$-graphyne, (b) $N_{ring}C$-graphyne, (c) $N_{chain}C$-graphyne, and (d) $(BN)_{64}$ structure, respectively. (The part circled by red line is the primitive cell. The carbon atoms are denoted by the grey dots, the nitrogen atoms, the blue dots, and the boron atoms, the pink dots, respectively.)

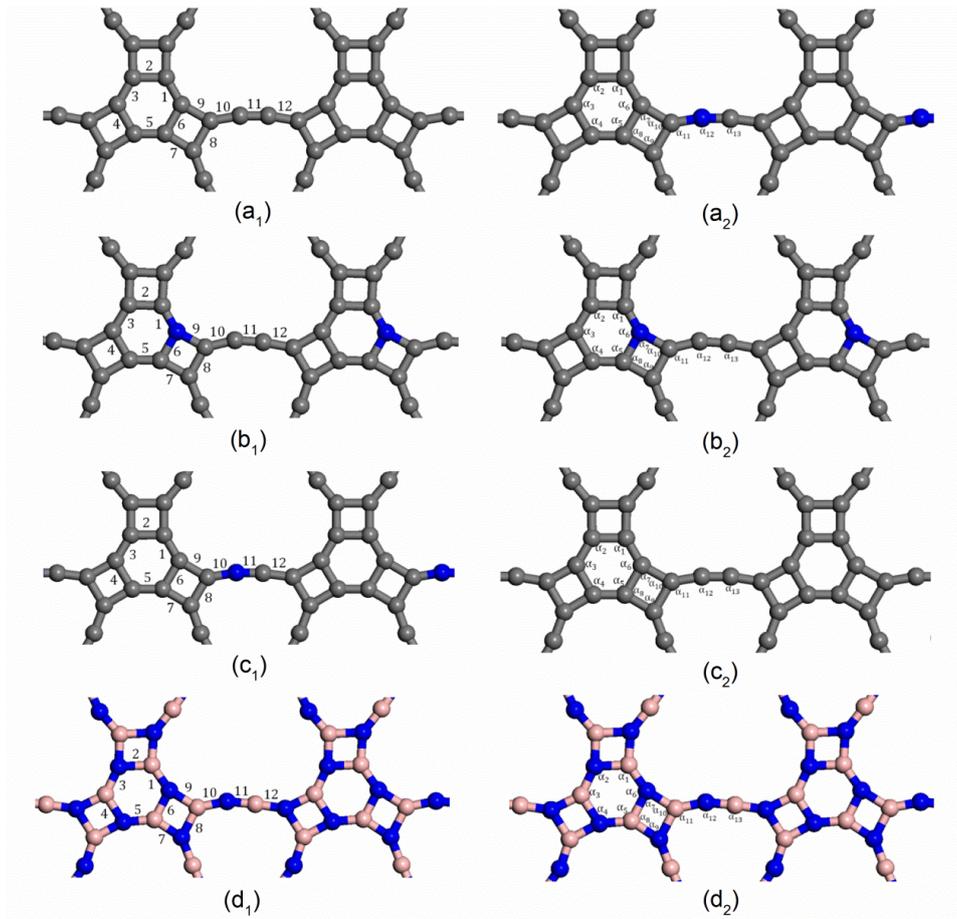

**Fig. 2.** The position mark of bonds and bond angles. ($a_1$, $a_2$) The bonds and bond angles of $C_{64}$-graphyne, ($b_1$, $b_2$) $N_{ring}C$-graphyne, ($c_1$, $c_2$) $N_{chain}C$-graphyne, and ($d_1$, $d_2$) $(BN)_{64}$ structure. The indexes 1–12 and $\alpha_1$-$\alpha_{13}$ denote the locations of various bonds and bond angles, respectively.

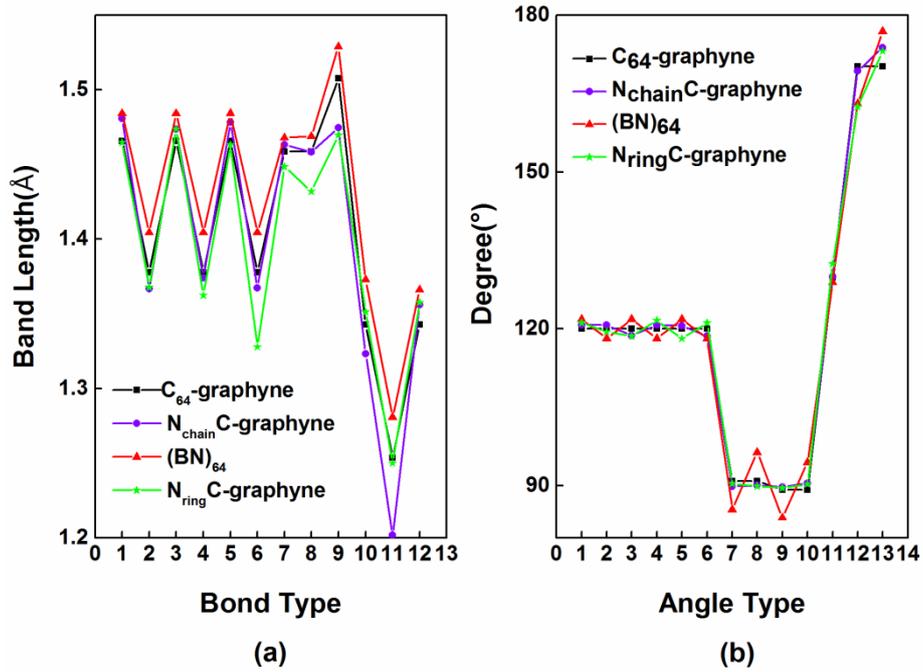

**Fig. 3.** The bond lengths and band angles of $C_{64}$-graphyne and its analogues. (a) Bond lengths, (b) bond angles. The black square, purple circle, red triangle and green pentagram represent the bond lengths and bond angles in $C_{64}$-graphyne, $N_{chain}$C-graphyne, $(BN)_{64}$ structure, and $N_{ring}$C-graphyne, respectively.

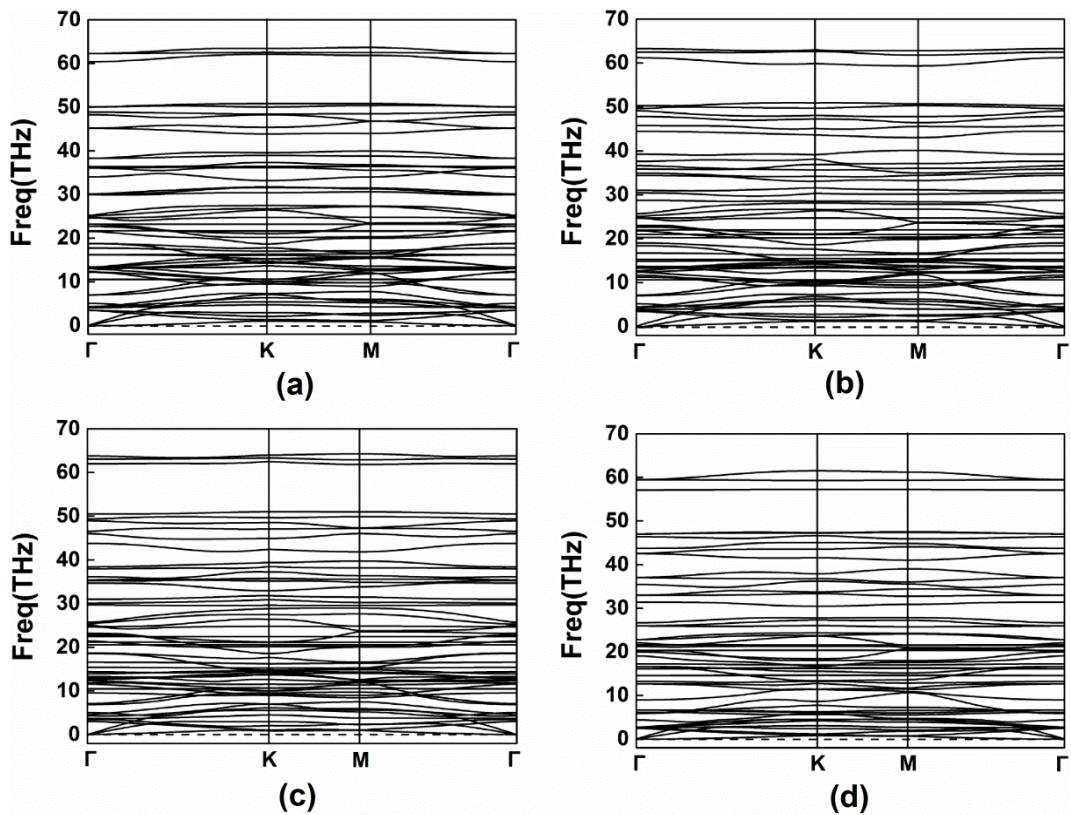

**Fig. 4.** Phonon spectra of (a) $C_{64}$-graphyne, (b) $N_{ring}$C-graphyne, (c) $N_{chain}$C-graphyne, and (d) $(BN)_{64}$ structure.

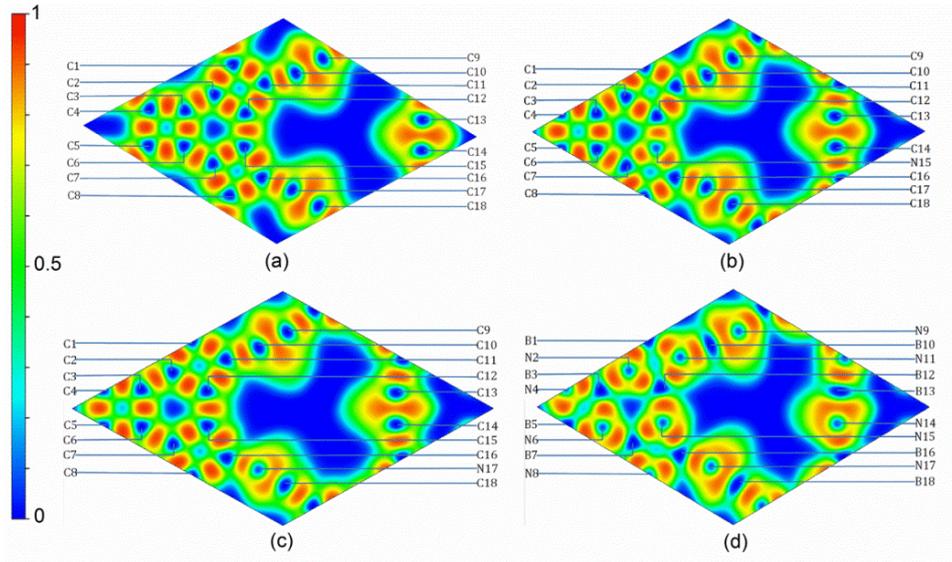

**Fig. 5.** The electron localization functions of (a) $C_{64}$-graphyne, (b) $N_{ring}C$-graphyne, (c) $N_{chain}C$-graphyne, and (d) $(BN)_{64}$ structure.

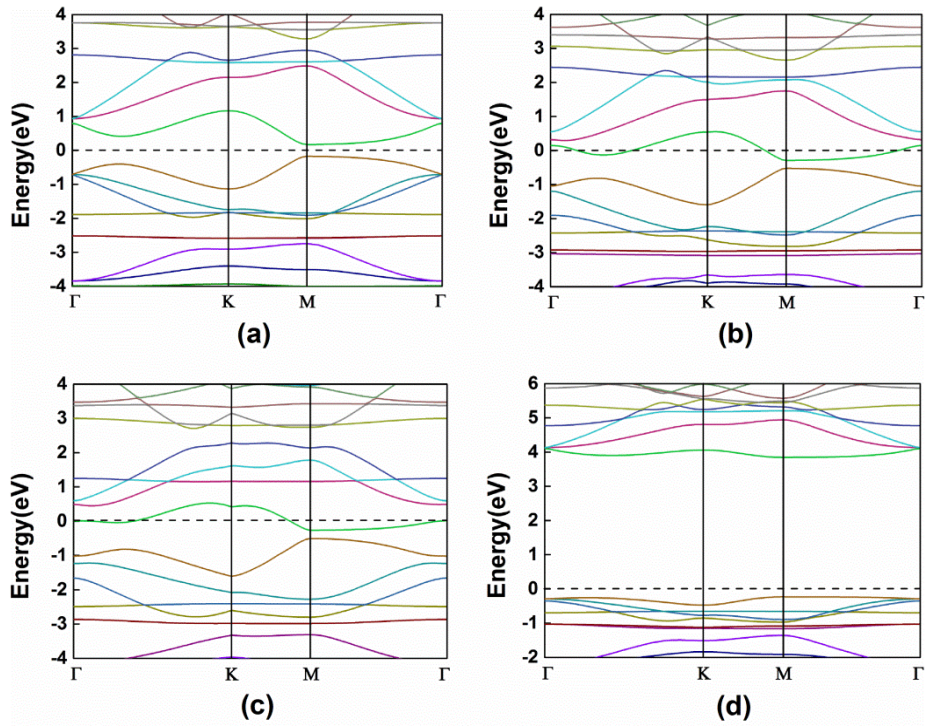

**Fig. 6.** The band structures of (a) $C_{64}$-graphyne, (b) $N_{ring}C$-graphyne, (c) $N_{chain}C$-graphyne, and (d) $(BN)_{64}$ structure.

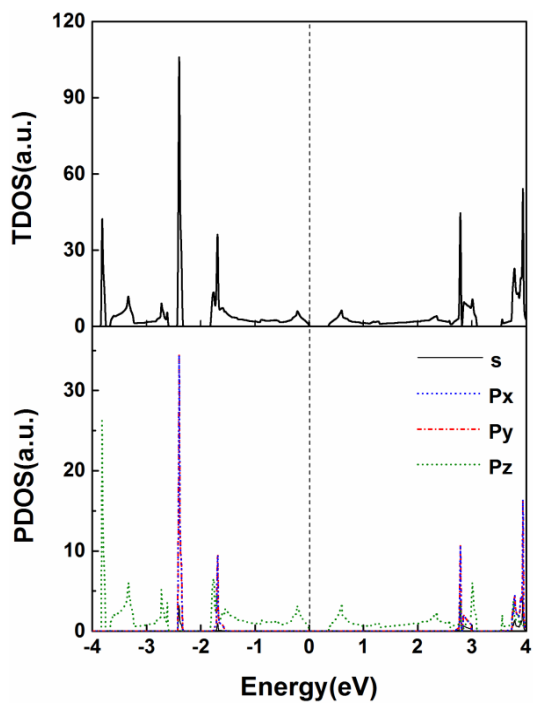

**Fig. 7.** The total and local density of states of $C_{64}$-graphyne.

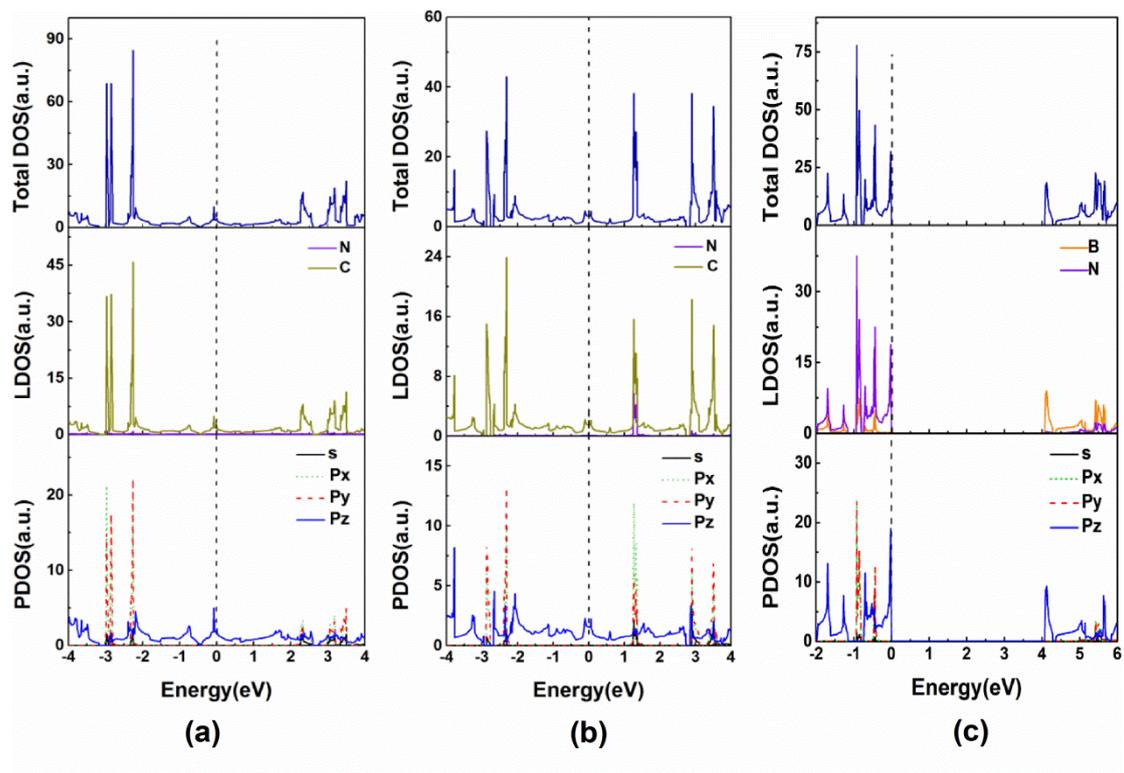

**Fig. 8.** The densities of states of (a) $N_{ring}C$-graphyne, (b) $N_{chain}C$-graphyne, and (c) $(BN)_{64}$ stucture (The Total DOS, LDOS and PDOS are from the top to the bottom).